\documentclass[a4paper,fleqn,usenatbib]{mnras}
\usepackage[T1]{fontenc}
\usepackage{epstopdf} 
\usepackage{graphicx}
\usepackage{float}
\usepackage{natbib}
\usepackage{hyperref}
\usepackage[english]{babel}
\usepackage{color}
\usepackage{todonotes}
\usepackage{epsfig}
\usepackage{epsf,latexsym}
\usepackage{amsmath}
\usepackage{amssymb}
\usepackage{caption}
\usepackage{subfig}
\epsfverbosetrue\usepackage{natbib}
\title[No evidence for large-scale outflows in Mrk273]{No evidence for large-scale outflows in the extended ionised halo of ULIRG Mrk273}
\author[R. A. W. Spence et al.]
	{R. A. W. Spence, $^{1}$\thanks{Email: rspence1@sheffield.ac.uk}
	J. Rodr\'{i}guez Zaur\'{i}n, $^{2}$
	C.  N. Tadhunter, $^{1}$
	M. Rose, $^{1}$
	A. Cabrera-Lavers, $^{2}$
	 \newauthor
	H. Spoon, $^{3}$
	C. Mu\~{n}oz-Tu\~{n}\'{o}n $^{2}$
\\
\\
$^{1}$Department of Physics \& Astronomy, University of Sheffield, Sheffield, S3 7RH, UK\\
$^{2}$Instituto de Astrofisica de Canarias (IAC), E-38205, La Laguna, Tenerife, Spain\\
$^{3}$Cornell Center for Astrophysics and Planetary Science, Space Sciences Building, Ithaca, NY 14853, USA
}

\date{Accepted XXX. Received YYY; in original form ZZZ}

\pubyear{2015}

\begin{document}
\label{firstpage}
\pagerange{\pageref{firstpage}--\pageref{lastpage}}
\maketitle

\begin{abstract}
We present deep new GTC/OSIRIS narrow-band images and optical WHT/ISIS long-slit spectroscopy of the merging system Mrk273 that show a spectacular extended halo of warm ionised gas out to a radius of $\sim45$ kpc from the system nucleus. Outside of the immediate nuclear regions (r > 6 kpc), there is no evidence for kinematic disturbance in the ionised gas: in the extended regions covered by our spectroscopic slits the emission lines are relatively narrow (FWHM $\lesssim$ 350 km$\rm s^{-1}$) and velocity shifts small (|$\Delta$V| $\lesssim{} $250 km$\rm s^{-1}$). This is despite the presence of powerful near-nuclear outflows (FWHM > 1000 km$\rm s^{-1}$; |$\Delta$V| > 400 km$\rm s^{-1}$; r < 6 kpc). Diagnostic ratio plots are fully consistent with Seyfert 2 photo-ionisation to the NE of the nuclear region, however to the SW the plots are more consistent with low-velocity radiative shock models. The kinematics of the ionised gas, combined with the fact that the main structures are aligned with low-surface-brightness tidal continuum features, are consistent with the idea that the ionised halo represents tidal debris left over from a possible triple-merger event, rather than a reservoir of outflowing gas.
\end{abstract}


\section{\label{sec:intro}Introduction}

Major gas-rich mergers are believed to be a key mechanism in driving galaxy evolution, where two (or multiple) galaxies coalesce, forcing gas and dust towards the nuclear regions and triggering both starburst and AGN activity. \citep[e.g.][]{DiMatteo2005}. The nuclear activity can create highly energetic outflows which blast material out of the galaxies, suppressing star formation and shaping the evolution of the systems. \par
It is well established that ultraluminous infrared galaxies (ULIRGs) in the local Universe are triggered close to the peaks of major mergers between gas-rich spiral galaxies. \citep[e.g.][]{Sanders1988, Veilleux2002}. Local ULIRGs are therefore ideal laboratories for detailed study, as they represent the phase in which nuclear outflows are expected to have the most significant effects on the host galaxy. Nuclear outflows have been detected in ULIRGs in all phases; from ionised \citep[e.g.][]{Rodriguez2013} to neutral \citep[e.g.][]{Rupke2011} to molecular \citep[e.g.][]{Cicone2014}, and it has been shown that the velocity widths (FWHM) and radial shifts of these outflows are substantially greater in ULIRGs showing signs of AGN activity \citep[e.g.][]{Rodriguez2013, Veilleux2013}. Previous observations, however, have focused on a radius of only a few kpc from the nucleus and evidence for larger-scale outflows is lacking, leading to uncertainties in quantifying the energetics. It is now important to determine the true spatial extents of these outflows and investigate the ultimate fate of the outflowing gas.  \par
In order to address these issues, we are undertaking a deep imaging and spectroscopic survey of a complete sample of 32 nearby ULIRGs. As part of this study we have observed Mrk273, one of the closest ULIRGs $(z = 0.0373, L_{IR} = 10^{12.21}L_{\sun})$ and a key target for detailed study. This object shows evidence for three nuclei: two detected at near- and mid-IR wavelengths and a third detected in [OIII] and radio emission (\cite{Rodriguez2014}; hereafter RZ14). The previous HST imaging and spectroscopic observations of RZ14 also shows a loop of [OIII] emission extending $\sim$20 kpc to the east of the nuclear region, as well as highly disturbed emission line kinematics within a radius of 6 kpc (see also \cite{Rupke2013}, RV13). The slits for these spectroscopic observations, however, did not sample the full extent of the halo, and the HST imaging did not probe the low-surface brightness features. The present study builds on this work and combines deep GTC/OSIRIS imaging with WHT/ISIS long-slit spectra in order to investigate the kinematics and ionisation mechanism of the extended halo of Mrk273. We adopt $H_{0}= 68 \rm km s^{-1}, \Omega_{m} = 0.29$ and $\Omega_{\Lambda} = 0.71$. The systemic redshift of the source gives a luminosity distance of $D_{L} = 171.5$ Mpc and a scale of 0.762 kpc $\rm arcsec^{-1}$.

\section{\label{sec:data}Observations and data reduction}

\subsection{\label{sec:imaging}Imaging data}
 H$\alpha$ Tunable filter (TF) imaging observations were taken for Mrk273 using the OSIRIS instrument\footnote{http://www.gtc.iac.es/en/pages/instrumentation/osiris.php} mounted on the 10.4m Gran Telescopio Canarias (GTC). To sample the H$\alpha$ emission from the galaxy, the TF was centered on 6810.7$\pm$2.0\AA{} (i.e. the red-shifted H$\alpha$ wavelength for the galaxy) and used with order sorter filter f680/43 ($\rm\lambda_{central}$ = 6802.1\AA{}; FWHM = 432\AA{}). These narrow-band observations had an effective bandpass of 17.7\AA{}, sensitive to $|\Delta\rm V_{H\alpha}| < ~400kms^{-1}$. In addition, a medium-band continuum image blueward of H$\alpha$ was taken using the OSIRIS red Order Sorter (OS) Filter f657/35 ($\rm\lambda_{central}$ = 6572\AA{}; FWHM = 350\AA{}). The exposure times were 1680 and 270 seconds for the H$\alpha$ and the continuum images respectively. \par
The TF and OS filter OSIRIS images were bias and flat-field corrected in the usual way. For the flux calibration, a photometric standard star was observed using exactly the same set up as for the galaxy. A flux calibration factor was then derived using a customised $\it python$ routine that generates a TF transmission curve for a given central wavelength and FWHM. \par
Any night sky lines contained within the bandwidth of the filter cause ring structures in the TF images. To remove such sky emission, a series of 1D polynomials were fitted in the x- and y-directions which generated a synthetic image containing only the sky emission. This image was then subtracted from the original, flux calibrated image of the source resulting in an image cleaned from any background and sky emission with the faintest visible structures having a surface brightness of $\sim 1.1\times10^{-17} \rm erg cm^{-1} s^{-1}arcsec^{-2}$   Full details on this process will be provided in Spence et al. (2016, in preparation).

\subsection{\label{sec:spec}Spectroscopic data}
Follow-up long-slit spectroscopic observations were taken in June 2014 using the Intermediate dispersion Spectrograph and Imaging System (ISIS) on the WHT telescope at the Observatorio del Roque de los Muchachos (ORM), La Palma. The 1.5" slit was offset by 2.7$\pm$0.1" (2.1$\pm$0.1 kpc) to the east of the continuum centre and positioned at PA 23 in order to include the extended emission to the NE of the nucleus and to the west of the tidal tail. The slit position is indicated in Figure \ref{fig:images}d. The observations were obtained using a dichroic cutting at 6100\AA{}, with R316R and R300B gratings. This instrumental setup resulted in a spatial scale of 0.4 arcsec pix$^{-1}$, and a spectral resolution of 5.4$\pm$0.1\AA{} in the blue and 5.0$\pm$0.1\AA{} in the red, as calculated using the mean widths (FWHM) of the night-sky lines. The data were reduced (bias subtracted, flat field corrected, cleaned of cosmic rays, wavelength calibrated and then flux calibrated) using the standard packages in IRAF and FIGARO. The relative flux calibration was estimated to be accurate to within 5\%.  The seeing for the observations was estimated to be $1.2 \pm 0.2 $ arcsec (FWHM). \par

\section{\label{sec:results}Results}

\subsection{\label{sec:morph}Morphology}

\begin{figure*}    
\begin{minipage}[t]{0.49\textwidth}
\centering
\subfloat[]{\includegraphics[width=\linewidth]{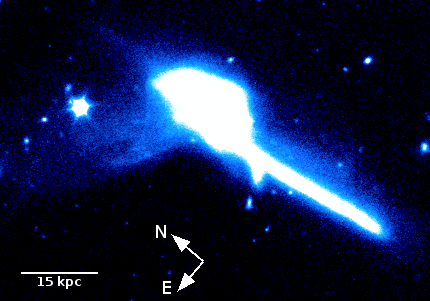}}
\label{fig:cont}
\end{minipage}
\hspace{0.1cm}
\begin{minipage}[t]{0.49\textwidth}
\centering
\subfloat[]{\includegraphics[width=\linewidth]{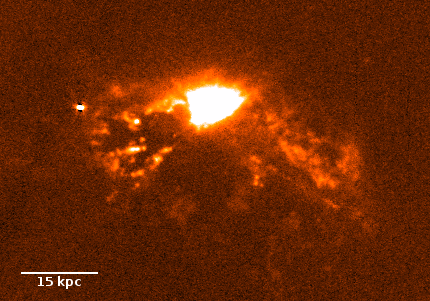}}
\label{fig:ha}
\end{minipage}

\vspace*{-0.05cm} 
\begin{minipage}[t]{0.49\textwidth}
\vspace{0.2cm}
\centering
\subfloat[]{\includegraphics[width=\linewidth]{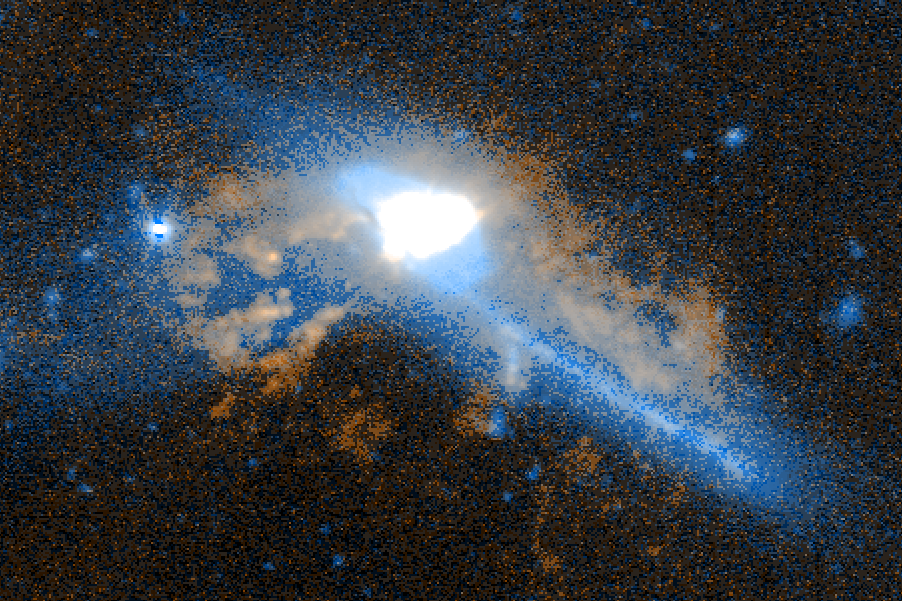}}
\label{fig:color}
\end{minipage}
\hspace{0.0cm}
\begin{minipage}[t]{0.49\textwidth}
\centering
\subfloat[]{\includegraphics[width=\linewidth]{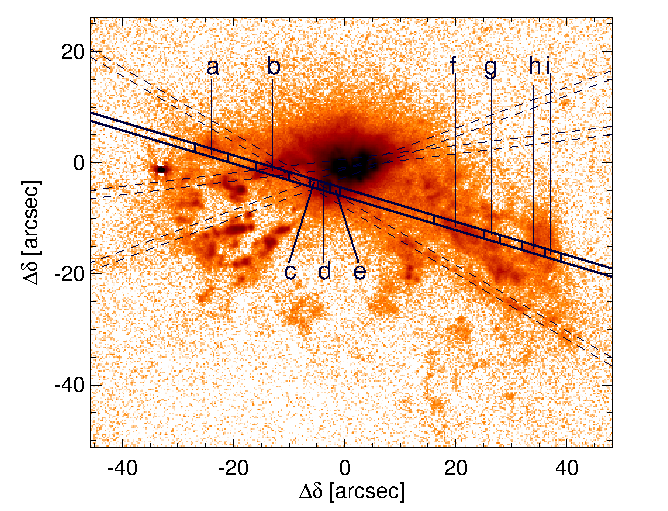}}
\label{fig:slits}
\end{minipage}
\vspace{-0.15cm}
\caption{Images of Mrk273 obtained with GTC/OSIRIS. \textbf{a)} Line-free continuum image showing the main tidal tail to the S and lower surface brightness diffuse emission to the NE and SW of the nuclear region. Note the extended loop to the NE. \textbf{b)} Continuum-subtracted H$\alpha$ image, showing the bright knots to the NE.  Also visible is the large amount of extended emission to the W of the tail. \textbf{c)} A composite colour image of the system. Blue and red/orange colours represent the continuum and H$\alpha$ emission respectively. \textbf{d)} H$\alpha$ image overplotted with the WHT/ISIS slit (PA 23) indicating the main apertures used for this paper (solid), along with the locations of our other spectroscopic slits (PA 56, 70 and 180, dashed; Rodriguez Zaur$\'{i}$n et al, 2010 and RZ14). Two further slits were shifted 1" to the N at PA 56 and 70, but are not plotted for clarity.}
\label{fig:images}

\end{figure*}

Our deep line-free continuum image of Mrk273 is presented in Figure \ref{fig:images}a. This shows the impressive tidal tail to the south of the galaxy, which represents debris from the main merger. The presence of a narrow dust lane in this tidal tail (RZ14) suggests that the disk of the main merging galaxy -- and the plane of the main merger -- is viewed close to edge-on. Notable is the appearance of an extensive looped continuum structure extending > 45 kpc to the east of the nucleus, at an angle of $\sim$130$^{\rm o}$ to the tidal tail. Assuming we are viewing the plane of the main merger edge-on, this structure could represent the tidal debris of a second merging galaxy which is approaching the nucleus at a large angle to the plane of the main merger. The conclusion that Mrk273 is a triple merger is further supported by the presence of three nuclei in the nuclear region (RZ14). We also find significant continuum emission to the west of the main tidal tail, but much less emission to the east. \par
Our deep continuum-subtracted H$\alpha$ image is presented in Figure \ref{fig:images}b. This shows a spectacular nebula of ionised gas, which stretches $\sim$ 40 kpc to the NE and SW of the nuclear region, with lower surface brightness regions also visible > 45 kpc to the SE. It is notable that the H$\alpha$ nebula to the NE stretches in a similar direction and with a similar spatial extent to the continuum loop. Note that the inner (< 20 kpc) higher surface brightness H$\alpha$ knots were also detected in [OIII] emission by RZ14. The substantial H$\alpha$ emission observed to the west of the main tidal tail is also striking, as little [OIII] emission was detected in this region in the previous HST imaging. This H$\alpha$ emission is most prominent to the west of the main tail, again following the continuum emission. The composite colour image in Figure \ref{fig:images}c highlights this.

\subsection{\label{sec:kin}Kinematics}
\begin{figure*}
	\centering
	\subfloat[]{\includegraphics[width=\columnwidth]{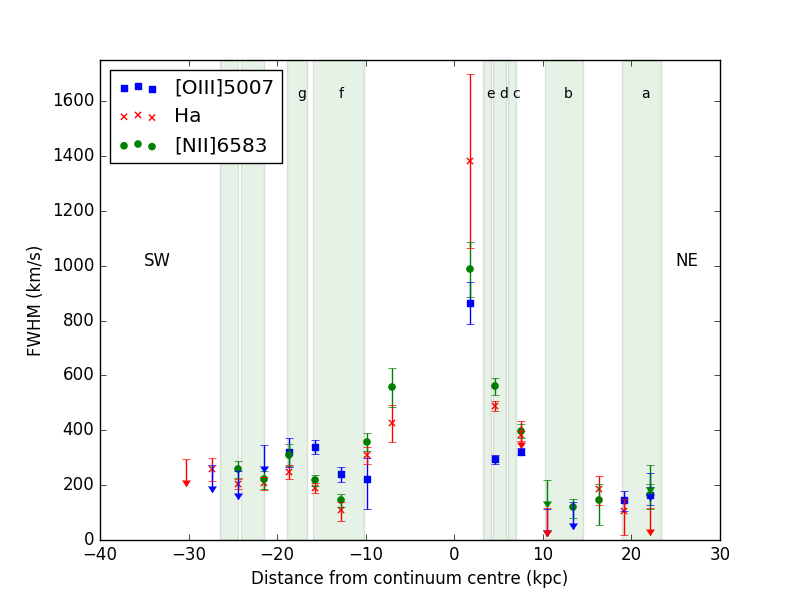}}
	\subfloat[]{\includegraphics[width=\columnwidth]{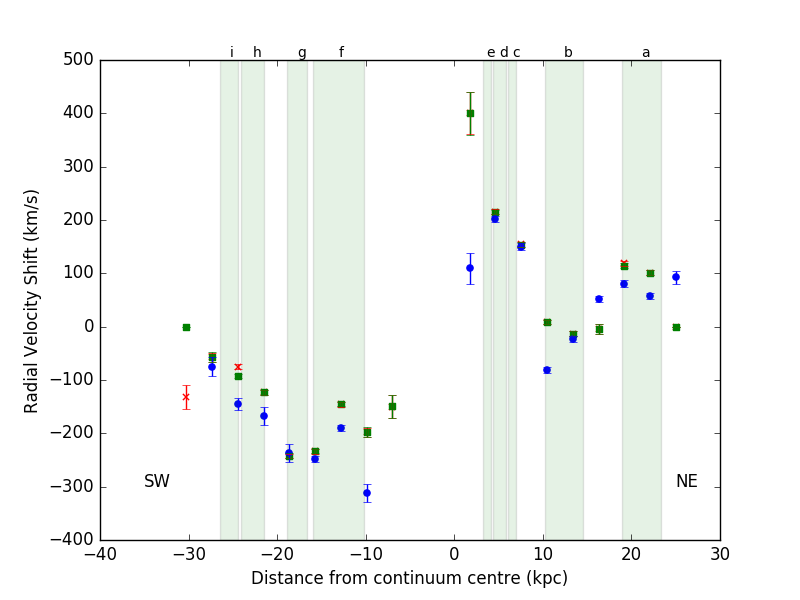}}
\vspace{-0.15cm}	
\caption{\textbf{a)} the velocity widths (FWHM) of the H$\alpha$, [NII]$\lambda$6583 and [OIII]$\lambda$5007 emission lines, corrected for instrumental width. Disregarding the aperture at $\sim$-8 kpc due to large error bars, quiescent gas (FWHM $\lesssim$ 350 km $\rm s^{-1}$) is found outside the range 0 < r < 5 kpc, corresponding to radial distance distance of $\sim$6 kpc from the nucleus. No complex kinematics are seen. The main apertures used to investigate the ionisation mechanism are indicated by the shaded areas. \textbf{b)} the velocity shifts relative to the continuum centre. Note there is an intrinsic uncertainty on these relative shifts due to the slit being offset from the system centre.} 
	\label{fig:kinematics}
\end{figure*}

Figure \ref{fig:kinematics} shows the velocity widths (FWHM) and radial velocity shifts ($\Delta V$) measured using single Gaussian fits at different locations along the slit. The reference point (D = 0 kpc) was chosen to be the centroid of the continuum emission along the slit, roughly corresponding to the position of closest approach to the nucleus. Broad lines (FWHM > 500 km$\rm s^{-1}$) are detected in the region where the slit intersects the near-nuclear region. These velocities are in agreement with previous studies \citep[e.g.][]{U2013, Rodriguez2014} and indicate the presence of powerful near-nuclear outflows. Apart from one region with large error bars at r $\sim$ -8 kpc, these broad lines appear to be confined to a narrow range along the slit (0 $\lesssim$ r $\lesssim$ 5 kpc), corresponding to a maximum radial distance of $\sim$6 kpc from the nucleus. \par
Outside the near-nuclear regions (r > 6 kpc), the H$\alpha$, [NII]$\lambda$6583 and [OIII]$\lambda$5007 emission lines are relatively narrow (FWHM $\lesssim$ 350 $\rm kms^{-1}$), and the velocity widths small (|$\Delta$ V| $\lesssim{}$ 250 km$\rm s^{-1}$). All emission lines are well fit with single Gaussians and show no signs of multiple kinematic components. We find the same quiescent kinematics in the extended regions in all our other slits, four from RZ14 and one from \cite{Rodriguez2009} (indicated on Figure \ref{fig:images}d). Despite extreme outflows with velocities up to 1500 km$\rm s^{-1}$ in the near-nuclear regions, these observations provide no clear evidence of the outflow extending beyond a radius of $\sim$ 6 kpc. \par

\subsection{\label{sec:ionize}Ionisation mechanism}

\begin{figure*}
	\centering
	\includegraphics[width=\textwidth]{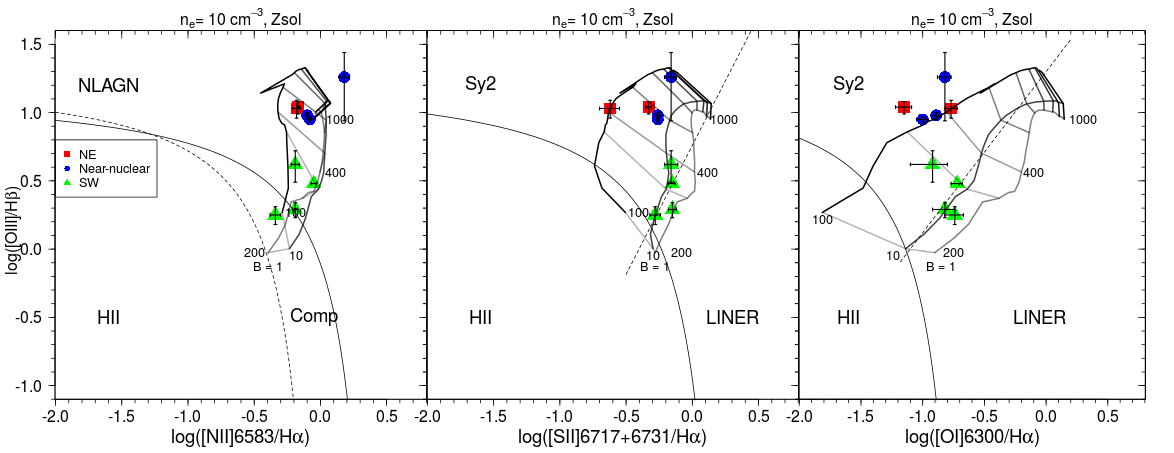}
\vspace{-0.6cm}	
\caption{BPT diagrams using emission line ratios as diagnostics for the dominant photo-ionisation mechanism. The overplotted dividing lines are described in section \ref{sec:ionize}. Also shown are the Allen et. al (2008) shock-ionisation model grids. The gridlines correspond to 3 values of magnetic field (B=1, 10 and 100 $\mu$G) and 9 shock velocities ($\rm V_{s} = 200, 300 ... 1000$ km$\rm s^{-1}$). $\rm V_{s}$ increases from bottom to top. The gridlines are grey-coded from light-grey to black, with light-grey corresponding to the lowest and highest values of B and $\rm V_{s}$. Notable is a clear divide between the northern and southern apertures suggesting different mechanisms dominate the two regions. The shock models are consistent with the SW ratios, but a poorer consistency is seen for the ratios in the NE.}.
	\label{fig:bpt}
\end{figure*}

The ionisation mechanism for the emission lines observed at different locations across Mrk273 was investigated using the Baldwin, Phillips \& Terlovich (BPT) diagnostic diagrams \citep{Baldwin1981, Veilleux1987}. The diagrams are shown in Figure \ref{fig:bpt}. Over-plotted on the first diagram are the theoretical "maximum starburst line" derived by \cite{Kewley2001} as an upper limit for star-forming galaxies (solid), and the semi-empirical lower boundary for star-forming galaxies derived by \cite{Kauffmann2003} (dashed). The area between these lines defines the composite region. Over-plotted on the second and third diagrams are the main AGN and Seyfert2/LINER empirical dividing lines from \cite{Kewley2006}.  Also over-plotted over all three diagrams are the \cite{Allen2008} pre-cursor + shock model grids for solar abundance and pre-shock density of 10$ \rm cm^{-3}$. The grids range in magnetic field (1 < B < 100 $\mu$G) and in shock velocity (200 < v < 1000 km$\rm s^{-1}$). A total of 9 apertures across the slit (denoted a to i) were analysed, sampling the extended emission to the NE and SW of the nuclear region, as well as the near-nuclear region. Figure \ref{fig:images}d shows the location of these apertures. For clarity, apertures a and b will be denoted "NE", c to e as "Near-nuclear" and f to i as "SW". The apertures are also indicated on the plots in Figure \ref{fig:kinematics} for comparison.  \par
The line ratios measured for both the NE nebula and nuclear regions are fully consistent with AGN (NLAGN/Sy2) photo-ionisation in all three diagnostic plots. The agreement with the shock models for these regions is poorer, particularly for the [OI]/H$\alpha$ ratio. This agrees with RZ14 who found that all areas of the nuclear region and extended filaments to the NE covered by their observations were consistent with Sy2 photo-ionisation, and the shock models generally failed to reproduce the line ratios.  The case for Sy2 photo-ionisation in these regions is further supported by the detection of HeII$\lambda4686$ emission to the NE. A HeII$\lambda4686$/H$\beta$ ratio of 0.30$\pm$0.05 was measured in aperture b, consistent with the typical values for AGN (HeII$\lambda4686$/H$\beta$ > 0.2; \cite{Osterbrock2006}). \par 
The nebula to the SW was not covered by the spectroscopic observations presented in RZ14 and is covered here by apertures f through i. There is an obvious division in Figure \ref{fig:bpt} between the ratios observed in the NE and SW of the galaxy, with the positions of the SW apertures more consistent with composite/LINER photo-ionisation than Sy2. The shock models reproduce the line ratios measured to the SW consistently across all three diagnostic plots, with shock velocities in the range 200 - 400 km$\rm s^{-1}$. These shock velocities are also consistent with the velocity widths measured at this location (Figure \ref{fig:kinematics}a). In comparison, AGN photo-ionisation models \citep{Groves2004} do not reproduce the line ratios across all three plots with the same level of consistency. 


\subsection{\label{sec:phys}Electron density, temperature and gas mass}
Single Gaussians were fitted to the [SII]$\lambda\lambda$6717,6731 emission lines in the extended apertures (a,b,f,g,h and i) to determine the electron density, and an error-weighted mean ratio of [SII]$\lambda$(6717/6731) = 1.4$\pm$0.1 was obtained, consistent with the low density limit. This leads to a 3$\sigma$ upper limit estimate of the density across the halo of $n_{e} \leq 430\rm cm^{-3}$ using the nebular modeling package FIVEL \citep{DeRobertis1987}.
The detection of [OIII]$\lambda$4363 in aperture b also allowed an estimate of the electron temperature. The ratio of the [OIII]$\lambda\lambda$(4959+5007)/4363 emission lines is 85$\pm$20, resulting in an electron temperature of $T = (1.37^{+1.8}_{-0.9})\times10^{4}K$.\par
Reddening is important in the nuclear regions of Mrk273 due to the large amounts of gas and dust present, however in the extended regions of our slit the degree of reddening appears to be small. We found an error weighted mean for H$\alpha$/H$\beta$ across the extended apertures of 3.1$\pm$0.1, consistent with case B recombination for AGN \citep{Gaskell1984}. \par
A lower limit for the warm gas mass of the extended halo (r > 6 kpc) was estimated using the H$\alpha$ luminosity (L(H$\alpha$) = $1.02 \times 10^{41} \rm erg s^{-1}$), measured in the extended regions by integrating the flux over the whole system ($r_{\rm aperture}$ = 45 kpc) and subtracting that of the nuclear region ($r_{\rm aperture}$= 6 kpc). Combining this with the effective recombination coefficient, $\alpha^{H\alpha}_{eff} = 0.77\times 10^{-13}$, for case B gas at $10{^4}$K (see \cite{Pequignot1991}) and the upper limit on the density presented earlier leads to a lower limit on the total halo gas mass of $\rm M_{halo} \gtrsim 8 \times 10^{5} M_{\sun}$.

\section{\label{ref:summary}Discussion and conclusions}
New GTC/OSIRIS and WHT/ISIS observations have been used to investigate the kinematics and ionisation mechanism of the extended halo of warm ionised gas surrounding the nuclear regions of Mrk273. Mrk273 shows both extended structure and extreme nuclear kinematics, which makes it an excellent object for investigating the connection between AGN-induced outflows and the host galaxy. AGN feedback is expected to be important to the evolution of galaxies yet current observational evidence is lacking for the impact of these outflows outside of the nuclear regions, on scales greater than a few kpc. This raises the question: to what extent do the AGN-driven outflows truly influence the outer regions of their host galaxies, and what is the ultimate fate of the outflowing gas?\par 
			Studies have shown the presence of an ionised outflow out to a 6 kpc radius from the nuclear regions of Mrk273. Therefore, one possibility is that the ionised gas in the extended nebula represents a reservoir of gas expelled from past/current outflows. Outflowing gas is generally identified by large velocity shifts and broad widths (FWHM) in emission line profiles. The shifts measured for the extended halo within our slit, however, are small (|$\Delta$V $\lesssim$ 250 km$\rm s^{-1}$) and the line widths narrow (FWHM $\lesssim$ 350 km$\rm s^{-1}$) -- inconsistent with the values observed in the nuclear regions by RV13 and RZ14. This is only consistent with the idea that the warm halo represents the accumulated reservoir of the outflowing gas if the turbulence in the outflow is somehow dissipated outside the near-nuclear regions. On the other hand, the relatively quiescent kinematics of the extended gas are entirely consistent with gravitational motions, given that the estimated total stellar mass of Mrk273 ($7\times10^{11} M_{\sun}$; \cite{Rodriguez2010}) is similar to that of the Milky Way. \par
Furthermore, RV13 found a mass outflow rate of 10 $ \rm M_{\sun}yr^{-1}$ for the inner warm outflow (r < 4 kpc). Using their typical outflow velocity of 750 km$\rm s^{-1}$, it would take $\sim$60 Myr to reach the outer extent of the halo ($\sim$45 kpc). Therefore the AGN would have a lifetime at least this long if the entire halo represents a reservoir of outflowing gas. Over this time the accumulated mass would be $\sim6\times10^{8} M_{\sun}$ for a constant mass outflow rate of 10 $ \rm M_{\sun}yr^{-1}$. This is a factor of $\sim$750 greater than the lower limit on the warm gas mass calculated in section \ref{sec:phys}; therefore, in order for the two masses to be consistent, the warm gas would have to be extremely rarefied ($\rm n_{e} \lesssim 1 cm^{-3}$). \par 
The most likely origin for the ionised halo in Mrk273 is gas left over from the merger(s) which triggered the activity. This is supported by the continuum "loop" observed to the NE of the nuclear region, which is clearly a tidal feature. The H$\alpha$ emission to the NE has a similar extent and stretches in the same direction as the continuum loop, also indicative of tidal debris. \par
We have analysed the ionisation mechanism for the ionised gas using BPT diagrams and Allen et al. (2008) radiative shock models. The model predicts the line ratios using a pre-cursor+shock model, with a pre-shock density of $\rm n_{e} = 10 cm^{-3}$, consistent with a compression factor $\sim$10 - 40, assuming the electron density measured in section \ref{sec:phys} is post-shock. The line ratios measured to the NE and in the near-nuclear regions are fully consistent with AGN photo-ionisation on the BPT diagrams (Fig. \ref{fig:bpt}). The Allen et. al (2008) shock models, in comparison, are less successful at reproducing the ratios (see also RZ14). These results lead us to conclude that the gas to the NE is tidal debris which is being illuminated by the nuclear AGN. \par 
The situation in the SW is less clear, with the BPT diagrams indicating that AGN photo-ionisation is a less dominant mechanism with respect to the NE and near-nuclear region. The line ratios fall mainly in the composite/LINER region and  are also consistent with the Allen et. al (2008) low-velocity radiative shock model, demonstrating that radiative shocks have the potential to produce LINER-like emission. Such low-velocity shocks can occur from collisions between clouds of gas and dust as they settle down after disruption by the major merger. The predicted shock velocities are in agreement with the measured velocity widths, shown in Figure \ref{fig:kinematics}, for the SW apertures. In comparison, AGN photo-ionisation models fail to reproduce the measured line ratios with the same level of consistency. Moreover, similarly to the NE, the ionised gas is observed to have a similar extent and direction as the continuum emission to the west of the tidal tail. Together, these results support the conclusion that the ionised gas to the SW is tidal debris which is being ionised by low-velocity shocks caused by turbulence from the merger(s).\par
Note that, if the SW gas were a reservoir from past outflows then the outflows in this location would most likely to be driven by intense star formation or supernovae in the tidal tail, stimulated by the merger. However this would not explain the asymmetry of the gas, as there would be no reason for the outflows to be preferentially directed to the west of the tidal tail. A further consideration is the travel time for any outflowing gas to reach the extended region. Although no evidence is seen here for large scale outflows, perhaps in this particular ULIRG the central AGN have not been active long enough for the outflows to reach kpc scales.\par
Overall, this study demonstrates the need for further detailed observations of extended emission line gas in local ULIRGs in order to probe the true extent of AGN-induced nuclear outflows and discover their true impact on the evolution of their host galaxies. 

\section*{Acknowledgements}
RAWS thanks Patricia Bessiere and Emmanuel Bernhard for their assistance with IRAF and \textit{IDL}, and Paul Kerry for his IT support. CMT acknowledges Estallidos project (AYA2013-47742-C4-2-P), funded by the Spanish Ministerio de Economia e Competitividad (MINECO). We thank support astronomer Gabi Gomez for his assistance during the observations. Based on observations made with the Gran Telescopio Canarias (GTC), installed in the Spanish Observatorio del Roque de los Muchachos of the Instituto de Astrof\'{i}sica de Canarias on the island of La Palma, and the William Herschel Telescope (WHT) which is operated on La Palma by the Isaac Newton Group. We thank the anonymous referee for useful comments.

\bibliographystyle{mnras}
\bibliography{mybiblio2}

\bsp
\label{lastpage}
\end{document}